\documentclass[conference]{IEEEtran}
\IEEEoverridecommandlockouts
\usepackage{amsmath,amssymb,amsfonts}
\usepackage{algorithmic}
\usepackage{graphicx}
\usepackage{textcomp}
\usepackage{xcolor}
\usepackage{graphicx} 
\usepackage{dcolumn}  
\usepackage{bm}       
\usepackage{url} 

\usepackage{graphicx}
\usepackage{amsmath}
\usepackage{hyperref}
\usepackage[square, sort, numbers]{natbib}

\usepackage{cite}
\usepackage[nameinlink,capitalize]{cleveref}  

\begin{document}

\title{Quantum-Enhanced Support Vector Machine for Large-Scale Stellar Classification with GPU Acceleration
}

\author{
    \IEEEauthorblockN{
        Kuan-Cheng Chen\IEEEauthorrefmark{1}\IEEEauthorrefmark{2}\IEEEauthorrefmark{7},
        Xiaotian Xu\IEEEauthorrefmark{1}\IEEEauthorrefmark{2},
        Henry Makhanov\IEEEauthorrefmark{3}\IEEEauthorrefmark{4},
        Hui-Hsuan Chung\IEEEauthorrefmark{5},
        Chen-Yu Liu\IEEEauthorrefmark{6},
    }
    \IEEEauthorblockA{
        \IEEEauthorrefmark{1}Centre for Quantum Engineering, Science and Technology (QuEST), Imperial College London, SW7 2BX London, UK\\
        \IEEEauthorrefmark{2}Department of Materials, Imperial College London, SW7 2BX, London, UK
    }
    \IEEEauthorblockA{
        \IEEEauthorrefmark{3}Department of Computer Science, The University of Texas at Austin, Austin, TX 78712, USA
    }
     \IEEEauthorblockA{
        \IEEEauthorrefmark{4}qBraid Co., Chicago, IL 60615, USA
    }
    \IEEEauthorblockA{
        \IEEEauthorrefmark{5}Max Planck Institute for Radio Astronomy, auf dem H{\"u}gel 69, D-53121 Bonn, Germany
    }
    \IEEEauthorblockA{
        \IEEEauthorrefmark{6}Graduate Institute of Applied Physics, National Taiwan University, 10663 Taipei, Taiwan
    }

    \IEEEauthorblockA{
        \IEEEauthorrefmark{7}Email: kuan-cheng.chen17@ic.ac.uk
    }
}

\maketitle

\begin{abstract}
In this study, we introduce an innovative Quantum-enhanced Support Vector Machine (QSVM) approach for stellar classification, leveraging the power of quantum computing and GPU acceleration. Our QSVM algorithm significantly surpasses traditional methods such as K-Nearest Neighbors (KNN) and Logistic Regression (LR), particularly in handling complex binary and multi-class scenarios within the Harvard stellar classification system. The integration of quantum principles notably enhances classification accuracy, while GPU acceleration using the cuQuantum SDK ensures computational efficiency and scalability for large datasets in quantum simulators. This synergy not only accelerates the processing process but also improves the accuracy of classifying diverse stellar types, setting a new benchmark in astronomical data analysis. Our findings underscore the transformative potential of quantum machine learning in astronomical research, marking a significant leap forward in both precision and processing speed for stellar classification. This advancement has broader implications for astrophysical and related scientific fields.

\end{abstract}


\section{Introduction}
Stellar classification stands as a cornerstone of astronomical research, entailing the categorization of stars according to their spectral properties \citep{bailerjones2002automated}. The spectra and the photometry of stars can help us infer their chemical components and intrinsic properties, thereby enhancing our understanding of the stellar evolutionary process. With the advent of large observational surveys and enormous amounts of astronomical data, the application of utilizing big data to study astronomical objects has become essential\citep{recio2016stellar, sen2022astronomical}. Traditional machine learning techniques, such as Support Vector Machines (SVMs), have largely contributed to stellar classification by exploiting spectral data to characterize stars with commendable accuracy \citep{kuntzer2016stellar}. However, the burgeoning volume and complexity of astronomical datasets are stretching the limits of these conventional methods, presenting formidable challenges in their applicability and scalability.

In the face of such complexity, quantum computing emerges as a beacon of innovation, offering novel solutions to the constraints of classical computational paradigms. QSVMs are at the vanguard of this technological forefront, demonstrating the potential to transform stellar classification methodologies. These advanced quantum machine learning algorithms are theorized to possess superior computational capabilities, potentially leading to heightened accuracy and expedited processing times in specific classification challenges \citep{chen2020variational, liu2023reinforcement}.

This study is dedicated to a meticulous exploration of QSVMs within the realm of stellar classification. We endeavour to craft and evaluate cutting-edge quantum algorithms, meticulously benchmarking their efficacy against established traditional methods. Our research is driven by the ambition to substantiate the strategic advantages that quantum computing holds in astronomical applications \citep{chen2022quantum_cnn}. A key aspect of our approach involves leveraging cuQuantum \citep{bayraktar2023cuquantum}, NVIDIA's toolkit for accelerating quantum circuit simulations on GPUs. By utilizing cuQuantum, we aim to enhance the scalability and accessibility of quantum-inspired algorithms, making them more feasible in the current era of non-fault-tolerant quantum computing. This integration is pivotal in addressing computational bottlenecks and propelling the efficiency of quantum simulations to new heights.

The anticipated outcomes of this inquiry not only promise to propel the methodology of machine learning for stellar classification forward but may also resonate across a wider spectrum of astrophysical disciplines and allied scientific territories, such as galaxy-star-quasar classification\citep{viquar2019machine}. The successful implementation of cuQuantum in streamlining quantum simulations underscores a significant stride in making advanced quantum algorithms more practical and applicable in the realm of astronomical research.

The advent of Quantum Machine Learning (QML) within a distributed quantum computing architecture heralds a significant paradigm shift in astronomical research, particularly in the realm of stellar classification. This development, exemplified by the potential integration of Quantum Federated Learning (QFL) with QSVM across a network of quantum processors \citep{prest2023quantum}, represents a transformative approach to data processing and analysis. Leveraging distributed quantum systems, this architecture enables the creation of collaborative learning models from decentralized datasets on diverse quantum devices, significantly enhancing both efficiency and privacy in data handling. Augmented by the integration of QSVM and recent advancements in Quantum Phase Estimation (QPE) through variational quantum circuit (VQC) approximation \citep{liu2023learning}, this paradigm shift marks a significant progression in quantum machine learning and computational strategies. Far from being a mere theoretical concept, this integration is emerging as a reality that promises to scale up the capabilities of QML, offering unprecedented accuracy and processing power for stellar classification tasks. It represents a stride towards a synergistic fusion of quantum and classical computational techniques, paving the way for significant advancements not only in astronomy but also in other scientific fields where the management of large-scale, sensitive data is crucial. As we explore this exciting frontier, the role of distributed quantum computing becomes increasingly central, reflecting the evolving landscape of computational methods in scientific exploration.

\subsection{Types of Stars and Current Methods of Classification}
\begin{figure}[htpb]
    \centering
    \includegraphics[width=0.45\textwidth]{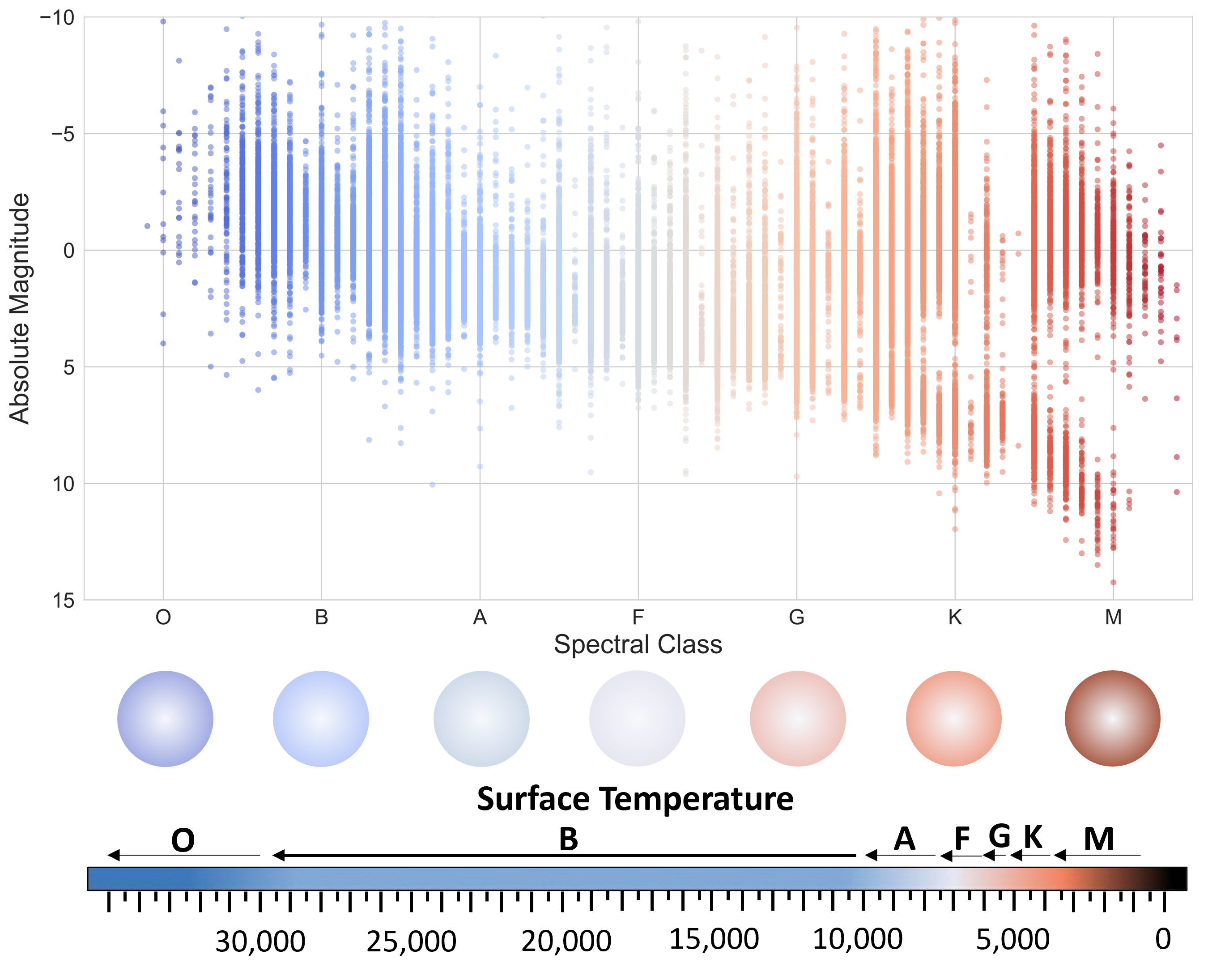}
    \caption{HR diagram highlighting the overlap in absolute magnitudes between spectral types A \& F and G \& K, illustrating the challenge in their distinction. The colour gradient reflects the temperature variation from hotter (blue) to cooler (red) stars.}
    \label{fig:star_class}
\end{figure}

The elucidation of stellar classifications hinges upon the analysis of stellar spectra, which provides insights into the temperature, luminosity, and chemical makeup of stars \citep{gray2021observation}. At the core of this scientific endeavour lies the Harvard spectral classification system, an enduring schema that categorizes stars into seven primary types (O, B, A, F, G, K, M), shown in Fig. \ref{fig:star_class}. These classifications draw from the seminal work of Annie Jump Cannon \citep{keenan1943atlas}, whose early 20th-century schematics continue to underpin modern astrophysics.

Spectral types serve as proxies for a star's thermal and luminous signatures, with O-type stars being the most incandescent and luminous, descending through to the cooler and dimmer M-type stars. Beyond mere temperature categorization, the luminosity class of a star offers a window into its dimensional attributes and stage of evolution, ranging from the expansive supergiants (Class I) to the more diminutive main sequence stars (Class V) \citep{cox2015allen}. The Hertzsprung-Russell (HR) diagram, an indispensable chart plotting stellar luminosity against surface temperature, further refines our understanding of these celestial bodies \citep{hrdiagram}. The HR diagram demarcates distinct domains occupied by stars at various junctures of mass, age, and evolution, offering a macroscopic view of stellar populations. Figure 1 in the present document exemplifies such HR diagram, assigning colours to stars based on their spectral classification, with the bluest hues representing the torrid O stars and the reddest shades corresponding to the temperate M stars.

Traditionally, the classification of stars has been a manual process, contingent on the meticulous examination of spectral lines, a practice that is not only laborious but also prone to inconsistency \citep{Lee2017}. The advent of machine learning, particularly the deployment of SVMs, has revolutionized this task by enabling a more automated and objective analysis, thereby mitigating human error \citep{Bailer_Jones2018}. Despite these advances, the burgeoning complexity and sheer volume of astronomical datasets are challenging the capabilities of conventional SVMs. It is within this context that quantum-enhanced SVMs emerge as a novel paradigm, promising to surpass the limitations of existing techniques and revolutionize the precision and computational efficiency of stellar classification \citep{Rebentrost2014}.

In this exploration, we delve into the potential of such quantum-augmented analytical methodologies to not only refine but also accelerate the classification processes that are pivotal to our comprehension of the stellar population.

\section{Methodology}

\subsection{Quantum Kernel Estimation Method}

Machine learning is a powerful tool for solving complex problems in fields ranging from computer vision to natural language processing. One of the most popular algorithms within this domain is the SVM, which is particularly effective for classification tasks. The advent of quantum computing has given rise to new possibilities in the field of machine learning, with QSVM being one of the most promising developments for classification problems.

\begin{figure}[htpb]
    \centering
    \includegraphics[width=0.45\textwidth]{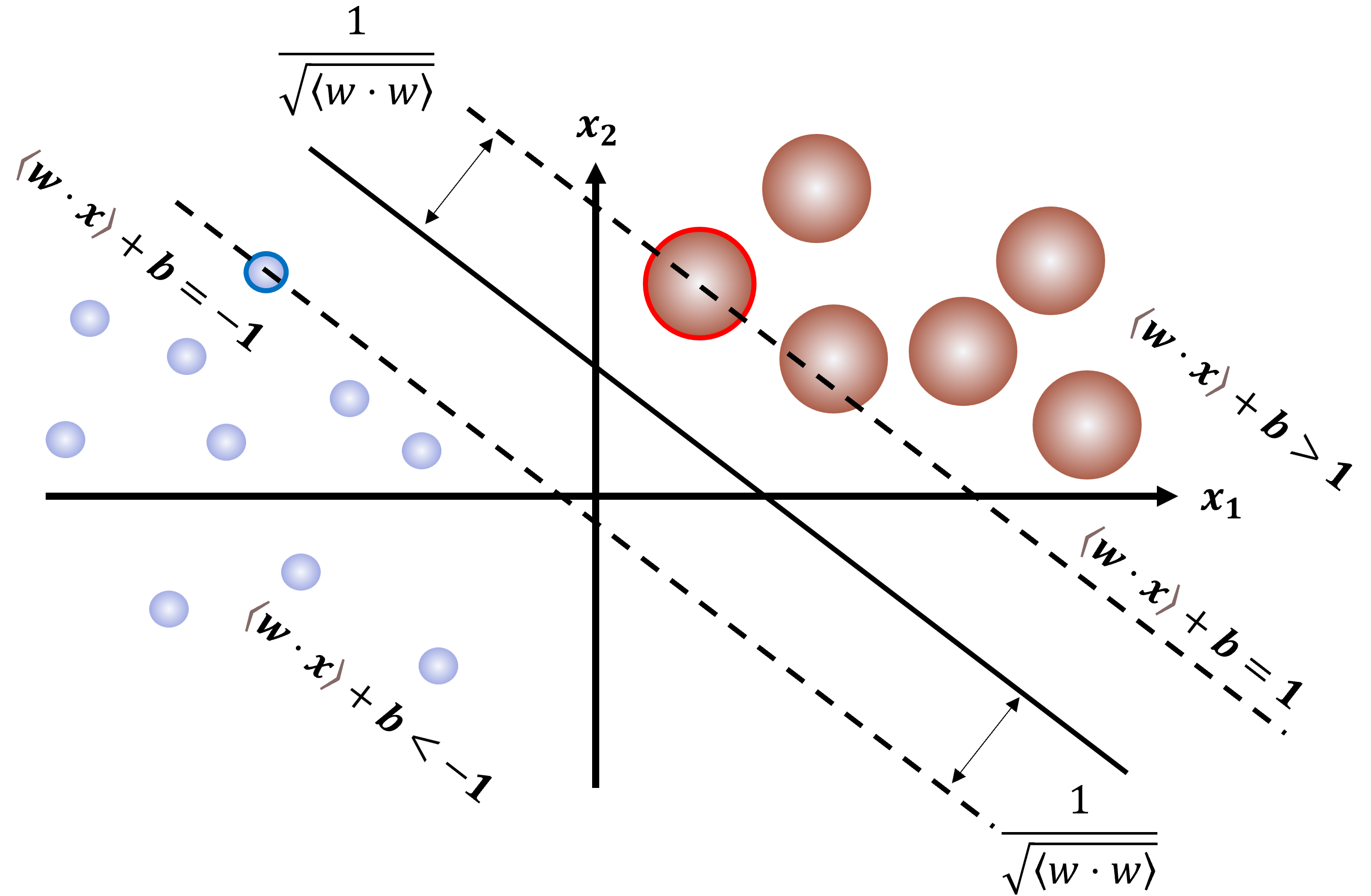}
    \caption{SVM classification of stellar data into dwarfs (blue) and giants (red), with the optimal hyperplane and margins distinguishing the two categories, and support vectors indicated.}
    \label{fig:svm}
\end{figure}

In classical SVM, the algorithm processes a set of input data points along with their respective labels to find the optimal hyperplane that separates the data into two distinct classes \citep{cortes1995support}. The objective is to maximize the margin between these classes, which in turn informs the classification of new data points based on their position relative to the hyperplane, as shown in Fig. \ref{fig:svm}.

QSVM extends the principles of the classical SVM into the quantum realm. The process involves utilizing a quantum kernel function that maps input data points to a high-dimensional Hilbert space. This mapping allows for the computation of the inner product between the data points, from which the probability of class affiliation is derived for classification purposes.

One common quantum kernel is the Gaussian kernel, implemented in the quantum setting as follows \citep{havlivcek2019supervised}:

\begin{equation}
    K(x_i, x_j) = \exp\left(-\frac{\|x_i - x_j\|^2}{2\sigma^2}\right),
\end{equation}

where \( x_i \) and \( x_j \) represent input data vectors and \( \sigma \) is the kernel's width parameter.

The principal advantage of QSVM over classical SVM is the utilization of quantum parallelism, an inherent attribute of quantum computing that enables the simultaneous computation of the kernel function across multiple data point pairs. This capability can theoretically lead to substantial computational speedups. For example, computing the kernel function for all \( n^2 \) pairs of data points could, in theory, be executed in \( O(\log n) \) time on a quantum computer, which is significantly faster compared to the \( O(n^2) \) time required on a classical computer\citep{gentinetta2022complexity}.

Additionally, QSVM is poised to exploit quantum entanglement to potentially surpass the classification performance of its classical counterpart. Entanglement, a unique quantum mechanical phenomenon, enables unprecedented correlations between particles that classical systems cannot replicate. By capitalizing on entanglement, QSVMs are expected to achieve greater classification accuracies, particularly beneficial in managing the large and intricate datasets commonly found in fields like astrophysics and high-energy physics\citep{wu2021application}.

\subsection{Star Dataset for QSVM Classification}

\begin{figure}[htpb]
    \centering
    \includegraphics[width=0.45\textwidth]{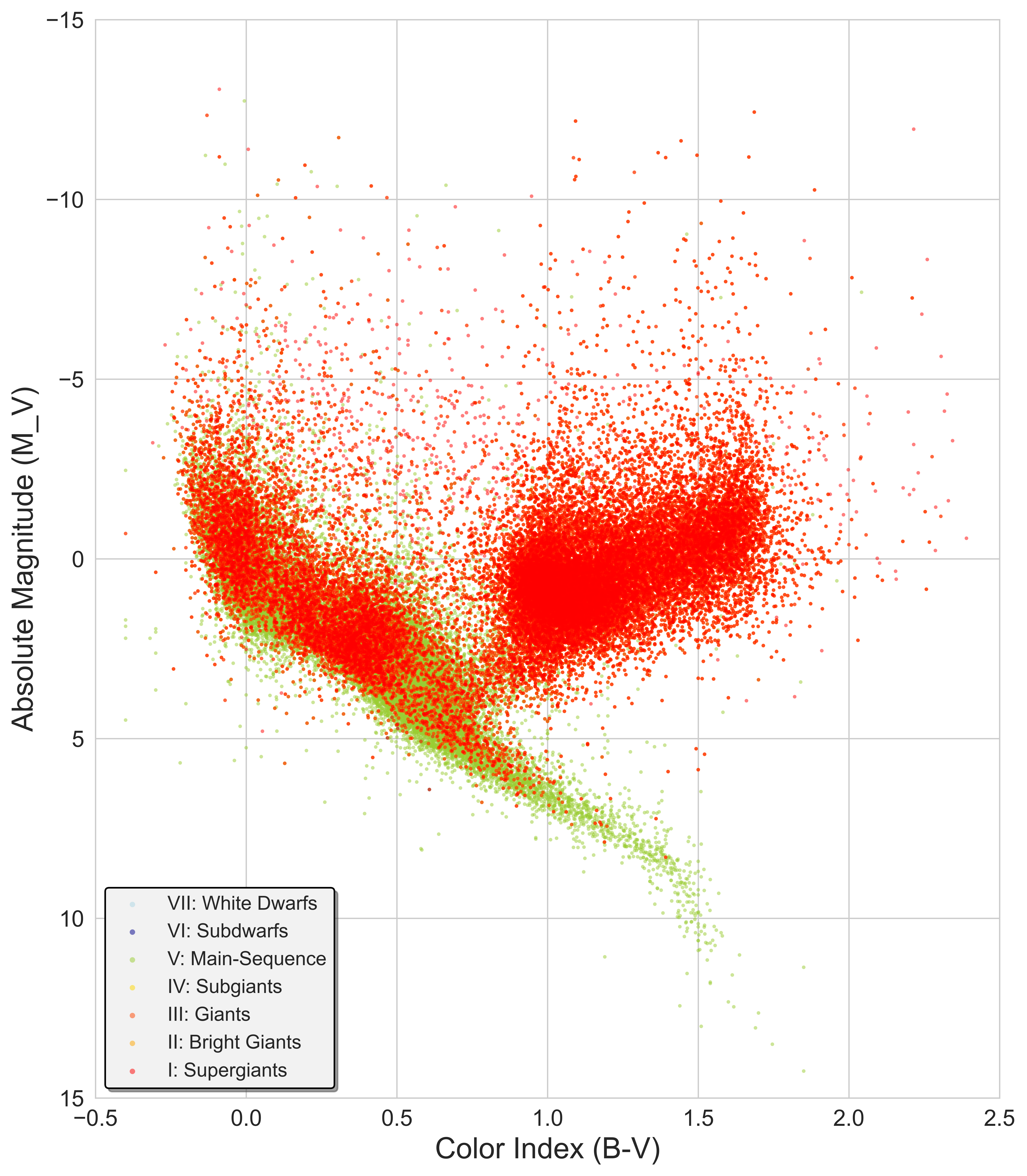}
    \caption{HR diagram from the Hipparcos Catalog, showing star classifications by luminosity and colour index, with distinct sequences for main sequence, giants, and white dwarfs.}
    \label{fig:star_dataset}
\end{figure}

The \textit{Star Categorization - Giants and Dwarfs} dataset, which is accessible on the Kaggle platform, presents a large collection of stars and their stellar properties based on multiple catalogues\citep{Ku_KaggleStellar, perryman1997hipparcos,ochsenbein2000vizier}. This collection encompasses data from more than 100,000 stars, enumerating attributes such as luminosity, temperature, radius, and mass. The relation among each category, colour index and absolute magnitude can be examined by the HR diagram shown in Fig. \ref{fig:star_dataset}.

\begin{figure*}[htpb]
    \centering
    \includegraphics[width=\textwidth]{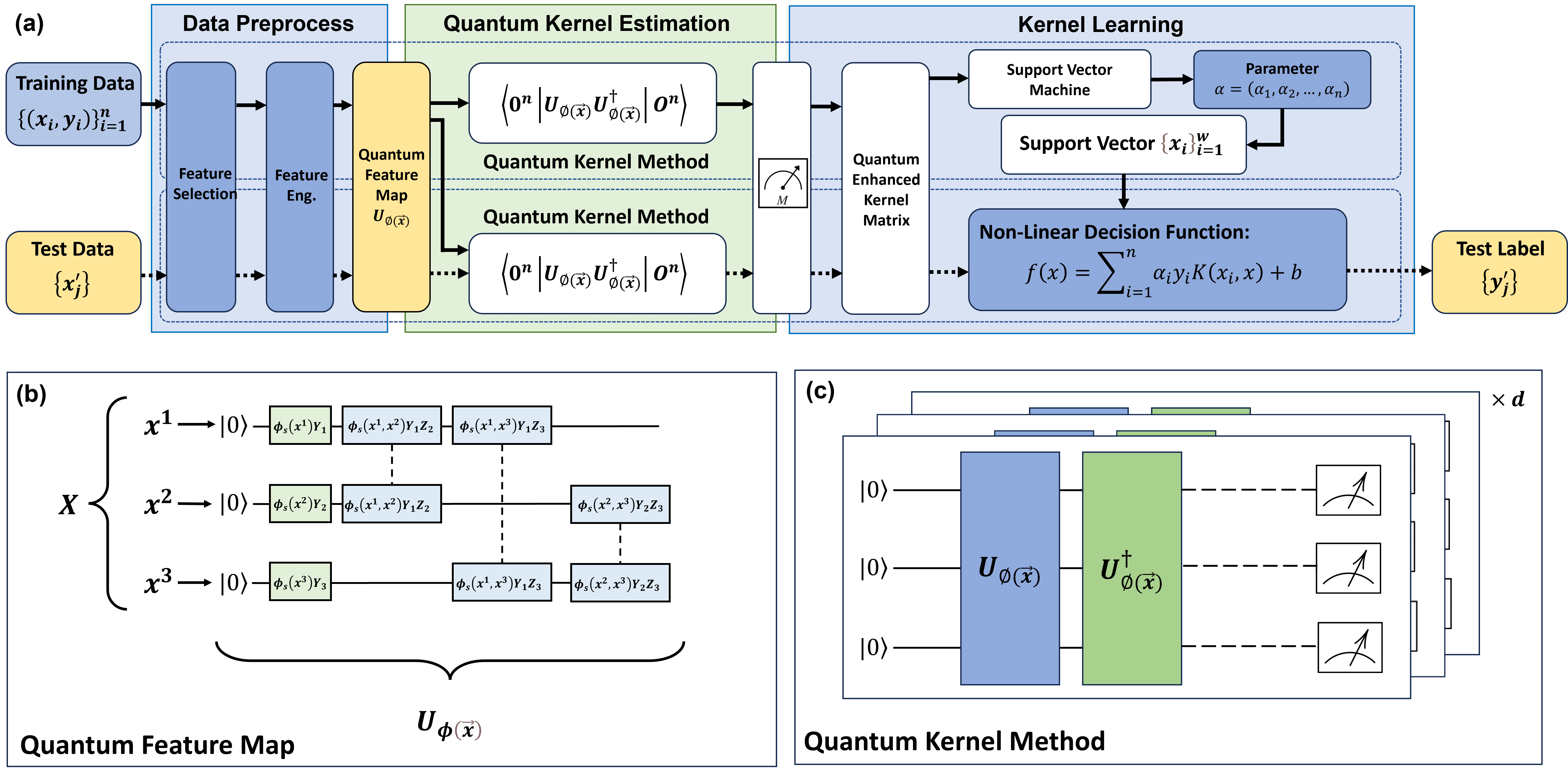}
    \caption{Overview of the Quantum Kernel Learning (QKL) algorithm: (a) depicts the process flow from data preprocessing to kernel learning, utilizing a quantum-enhanced kernel matrix and subsequent SVM for classification; (b) illustrates the quantum circuit representation of the quantum feature map used in transforming classical data into quantum states; and (c) demonstrates the quantum circuit for the quantum kernel method, highlighting the computation of inner products in the Hilbert space.\citep{havlivcek2019supervised}.}
    \label{fig:qsvm_circuit}
\end{figure*}

Within this dataset, stars are systematically classified into two predominant groups: giants and dwarfs. Giants are characterized as stars that surpass the Sun in terms of both luminosity and size, whereas dwarfs are smaller and less luminous when compared with the Sun. The classification is deduced from the values of luminosity and radius, which are themselves derived from spectral data. This dataset is essential in the study of stellar classification, which utilizes spectral data to categorize stars into various categories. Predominantly based on the Morgan–Keenan (MK) classification system\citep{morgan1973spectral}, it employs the historical HR classification system for chromaticity categorization and uses Roman numerals for sizing stars. Essential to this dataset is the absolute magnitude and B-V Color Index, which are pivotal in distinguishing giants and dwarfs, thus offering a comprehensive insight into the distinct characteristics of these celestial bodies.

In the context of this paper, the dataset serves as a foundational framework for examining various stellar attributes, as well as for the development of machine learning models that can effectively classify stars into giants or dwarfs according to their stellar parameters. Furthermore, it provides a benchmark for evaluating the efficacy of quantum machine learning algorithms within the domains of astronomy. This study using quantum machine learning would explore the possibility of utilizing the quantum algorithm in astronomy research in the generation of big data.

\subsubsection{Data Preprocessing for Astrophysical Data}

\begin{figure}[htpb]
    \centering
    \includegraphics[width=0.45\textwidth]{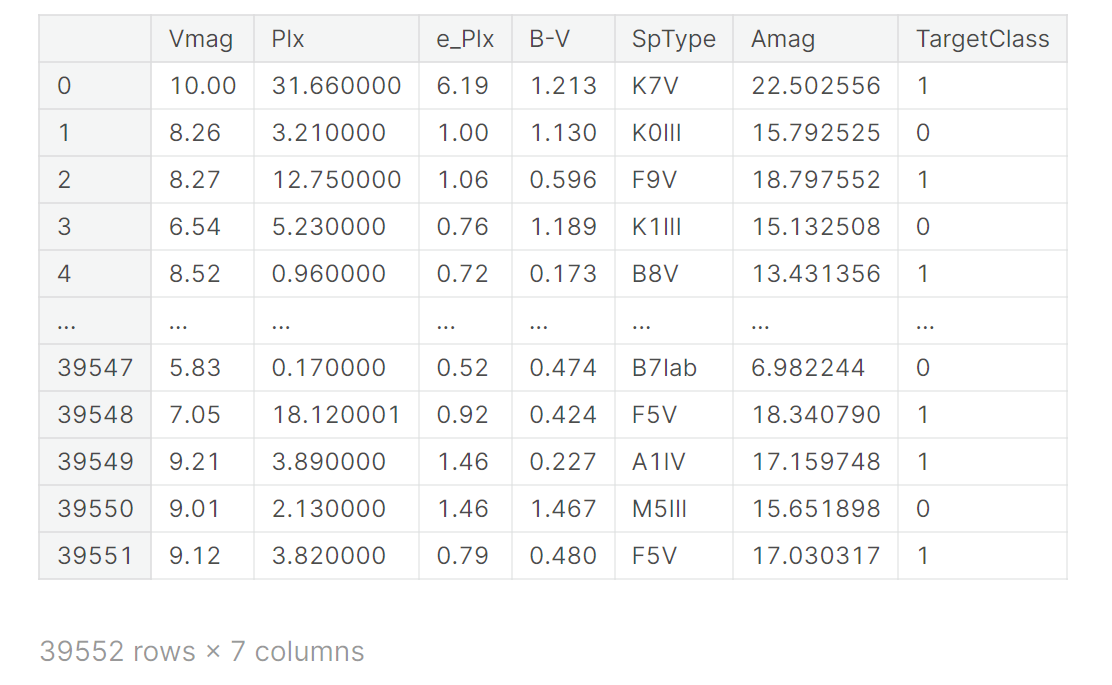}
    \caption{Post-preprocessing dataset snapshot, encompassing 39,552 data points across 5 parameters and 2 labels (SpType and TargetClass).}
    \label{fig:processed_data}
\end{figure}

Data preprocessing, shown in Fig. \ref{fig:processed_data}, is a critical step in machine learning, tailoring data for analysis and enhancing the efficacy of the algorithms applied. Within the scope of the \textit{ Star Categorization - Giants and Dwarfs} dataset, preprocessing encompasses a series of actions designed to clean and refine the data. The steps of preprocessing are stated in the following bullet points:

\begin{itemize}

     \item The initial phase involves the removal of duplicates and the rectification of missing values. This procedure aims to maintain the dataset's integrity and consistency, thereby avoiding discrepancies or errors.

     \item Subsequently, the dataset is subjected to standardization, aiming to neutralize any potential biases or variations arising from divergent scales among features. This process involves the normalization of feature scales by subtracting the mean and dividing by the standard deviation, thus yielding features with a mean of zero and a standard deviation of one. This equalizes the significance of all features during analysis and prevents any undue impact on the machine learning models due to the scale of features.

     \item Following the standardization, the dataset is divided into training and testing subsets. This division allows for the machine learning models to be trained on one subset of data and then validated for accuracy and generalizability on a distinct subset that has not been previously utilized in the training process.

     \item The final step in data preprocessing is encoding, which converts categorical data into a numerical format amenable to machine learning algorithms. A prevalent strategy for this conversion is one-hot encoding, which produces columns corresponding to each categorical variable in the dataset.

\end{itemize}

\section{Model}

In this study, we harness Quantum Kernel Learning (QKL) — a QSVM — to navigate the intricacies of stellar classification. The overview of the QKL algorithm is shown in Fig. \ref{fig:qsvm_circuit}.  QKL's advanced computational capabilities are leveraged to effectively and precisely categorize stars, particularly distinguishing between white dwarfs and giants. By utilizing the quantum realm's rich feature space, QKL is expected to deliver superior accuracy and performance over conventional classification methods. Throughout this paper, the term QKL will be used to denote this quantum-augmented machine-learning approach.

\subsection{Feature Engineering for Star-Classification Dataset}

For the task of classifying stars as either white dwarfs or giants, QKL emerges as a particularly promising technique, given its potential for high precision in predictions. Nevertheless, the intrinsic characteristics of quantum data, coupled with the current limitations in quantum hardware access, elevate the importance of feature engineering within this domain. To optimize data for QKL, traditional preprocessing methods are employed to streamline the dataset's dimensionality and to isolate pertinent features prior to engaging the quantum kernel. This strategy is aimed at diminishing the computational load and enhancing model accuracy by curtailing the noise and ambiguities inherent in quantum datasets.

One approach to readying the data for QKL involves the conversion of the quantum state into a representation more conducive to machine learning algorithms. This may entail the application of quantum circuits or alternative transformations to distil beneficial features. For instance, the B-V colour index, indicative of stellar temperature, could be encoded into a quantum state, given its significance in star classification. In a similar vein, the spectral type could be encoded to reflect the pattern of spectral lines within the star's spectrum, a crucial classification determinant.

The feature selection process exerts a considerable influence on the quantum kernel's accuracy, as the kernel's role is to assess the similarity between data point pairs. The features chosen for comparison, therefore, are pivotal. Meticulous feature engineering is imperative to ensure the kernel is attuned to the data's most informative attributes. For example, the visual apparent magnitude and absolute magnitude are likely to be critical for accurate star classification, while other attributes, such as \(e\_Plx\), may lack relevance and could be excluded. The essence of feature engineering, therefore, is paramount to the efficacy of QKL in stellar classification tasks.
\section{Feature Engineering for Stellar Classification}

Stellar classification leverages various attributes such as visual apparent magnitude (\verb|Vmag|), parallax (\verb|Plx|), B-V colour index (\verb|B-V|), spectral type (\verb|SpType|), and absolute magnitude (\verb|Amag|) to differentiate between white dwarfs and giants. Each of these attributes plays a distinctive role in the classification process:

\begin{itemize}
    \item \verb|Vmag|, or visual apparent magnitude, represents a star's apparent brightness observed from Earth. 
    \item \verb|Plx|, representing the star's parallax, is essential for calculating its distance. This measurement is critical in determining the star's absolute magnitude, thereby enabling the differentiation between the intrinsic luminosity of dwarfs and giants.
    \item \verb|B-V|, the colour index, is derived from the B (Blue) and V (Visual) passbands in photometric observations. A smaller \verb|B-V| value indicates a bluer and hotter star.
    \item \verb|SpType| reveals a star's physical properties through spectral lines, essential for categorizing stars into their respective classes.
    \item \verb|Amag|, or absolute magnitude, provides insight into a star's intrinsic brightness and size, thereby serving as a key discriminator between white dwarfs and giants.
\end{itemize}

Feature engineering in this context refines raw astronomical data for QKL techniques. This process includes standardizing numerical attributes, encoding categorical variables into quantum-compatible formats, and potentially developing composite parameters to better capture the intricacies of stellar data. Performing accurate classification through these refined features has far-reaching implications for interpreting stellar attributes and understanding stellar evolution.

The relationship between the attributes is captured in the following equation:

\begin{equation}
    \verb|Amag| - \verb|Vmag| = 5 \log_{10}(\verb|Plx|) + 5
\end{equation}

To introduce additional variance, composite parameters are defined as follows:

\begin{equation}
\begin{aligned}
    &\verb|Amag_SQ| = \verb|Amag|^2,\\
    &\verb|B-V_SQ| = \verb|B-V|^2,\\
    &\verb|B-V+Amag| = \verb|B-V| + \verb|Amag|,\\
    &\verb|B-V-Amag| = \verb|B-V| - \verb|Amag|.
\end{aligned}
\end{equation}

In our model, we hypothesize that \verb|Amag| (absolute magnitude) and \verb|B-V| (colour index) are the most critical features for distinguishing stellar types, as they represent the luminosity and temperature of stars, respectively—key parameters in the HR diagram. Conversely, \verb|Vmag| (visual magnitude) might be less critical in this context. Our classification approach will test this hypothesis using quantum-enhanced machine learning methods, evaluating how effectively these features can differentiate between various stellar classifications in accordance with their positions and trajectories on the HR diagram.

\subsection{Feature Interactions in Stellar Classification}

\begin{figure}[htpb]
    \centering
    \includegraphics[width=0.45\textwidth]{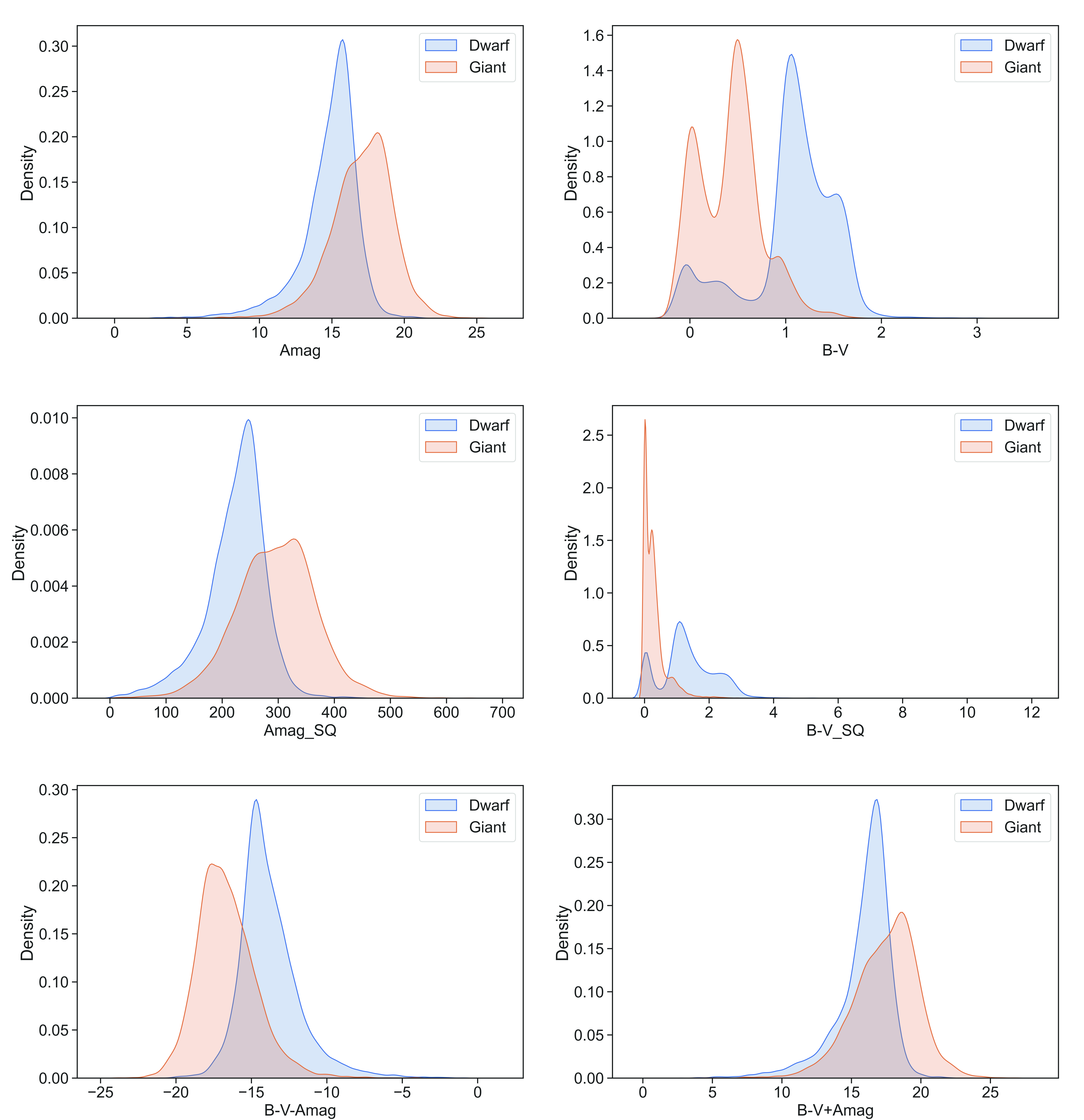}
    \caption{The scatter plot visualises the relationship between various features, including absolute magnitude, B-V colour index, and their composite transformations. This figure elucidates the dependencies and interactions of these characteristics, which are crucial in the categorization of stars}
    \label{fig:feature_relations}
\end{figure}

\begin{figure}[htpb]
    \centering
    \includegraphics[width=0.45\textwidth]{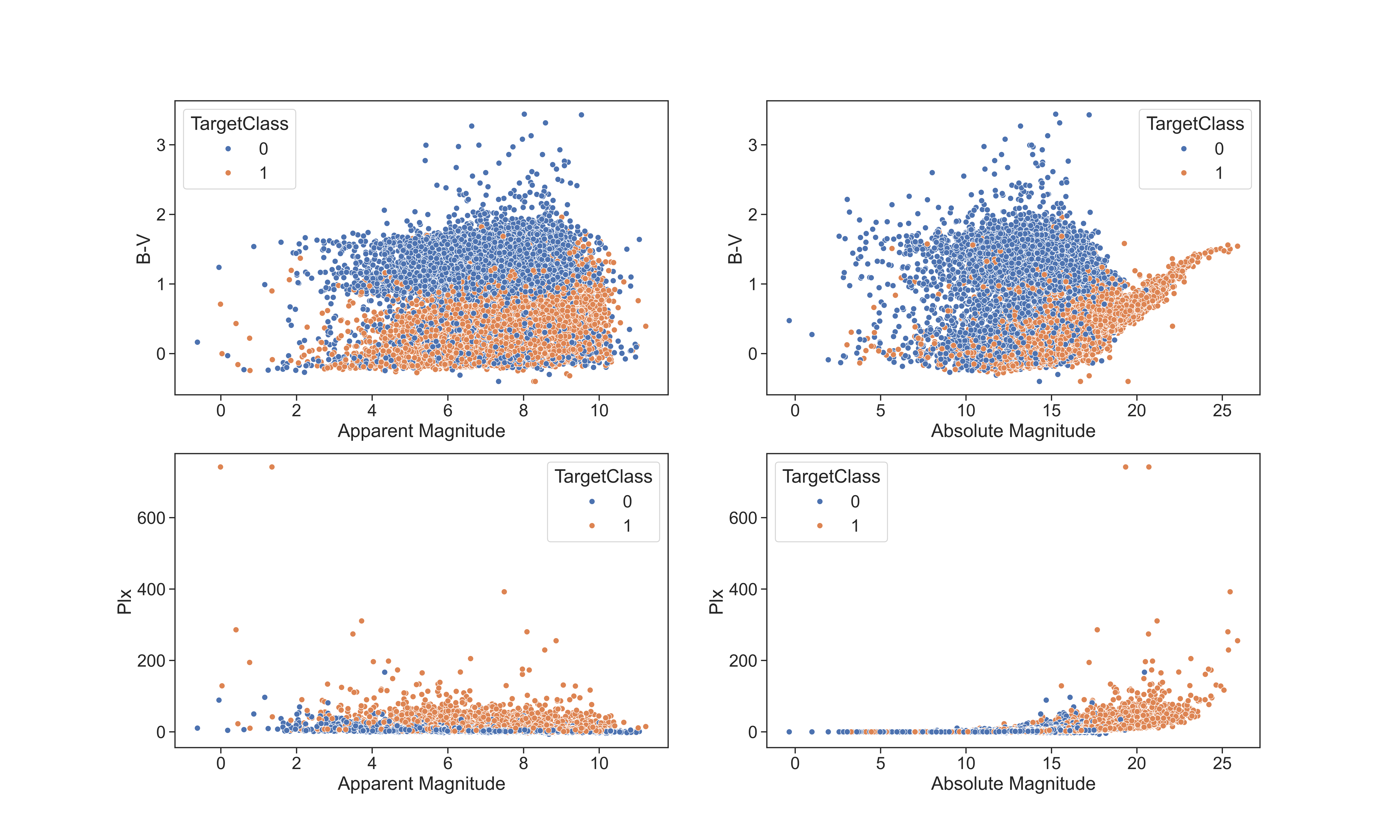}
    \caption{Interaction graph between absolute magnitude and the colour index as well as parallax, demonstrating feature behaviour for binary classification of stars (dwarf and giant).}
    \label{fig:data_2class}
\end{figure}

Understanding the intricate relationships among features is crucial for the classification. Graphical representations of these interactions provide insights into how different attributes, such as absolute magnitude and B-V colour index, influence the stellar classification (a dwarf or a giant), shown in Fig. \ref{fig:feature_relations} and Fig. \ref{fig:data_2class}.

Analysis of these figures reveals the efficacy of composite features in highlighting subtle variances across attributes, enhancing the capabilities for the classification. Particularly, such features facilitate a clearer separation between data clusters, thereby potentially increasing the accuracy of the classification.

\begin{figure}[htpb]
    \centering
    \includegraphics[width=0.45\textwidth]{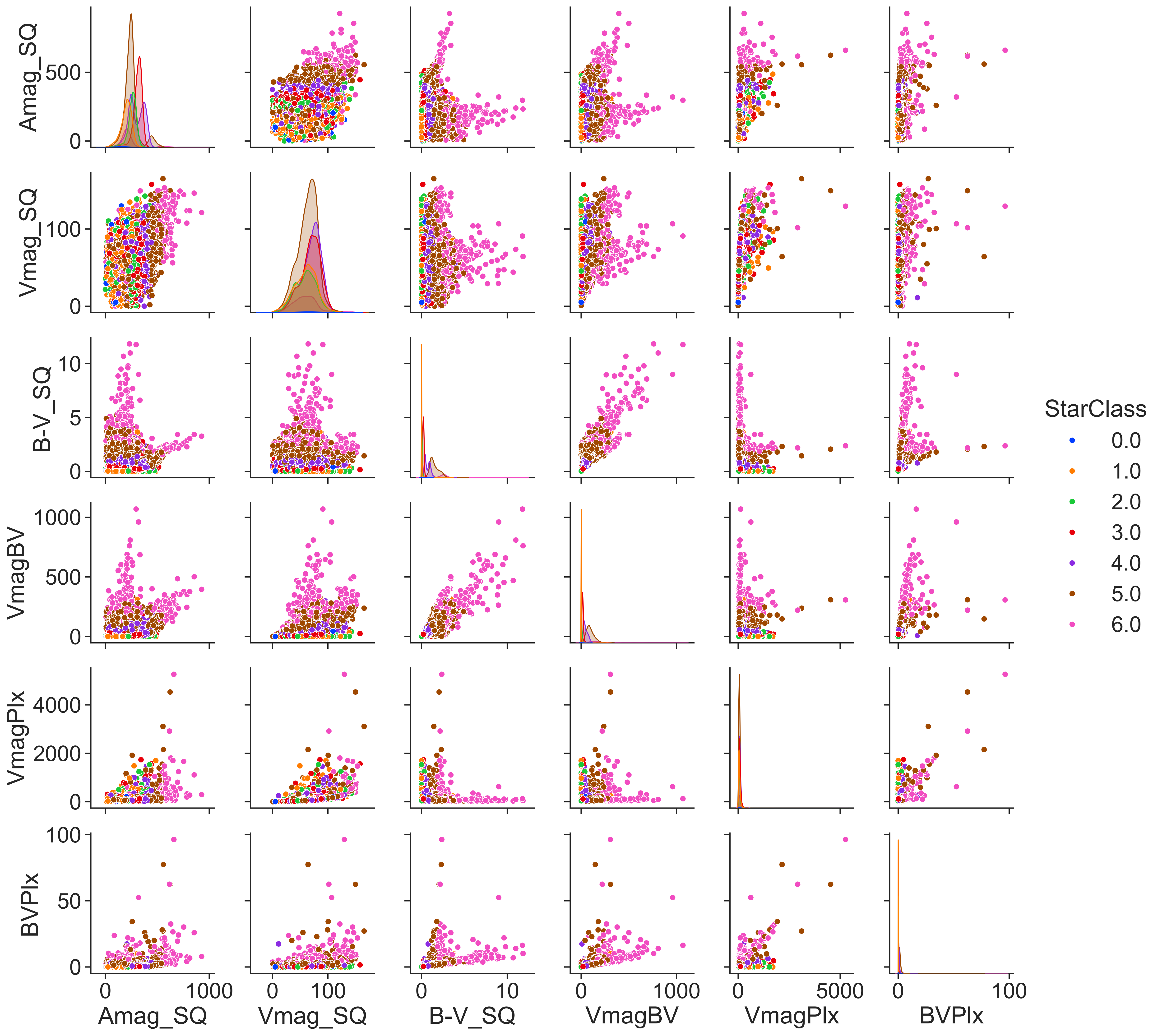}
    \caption{Detailed interaction graph for each pair of features within the dataset.}
    \label{fig:interactions_white}
\end{figure}

The use of such composite parameters, as demonstrated in  Fig. \ref{fig:interactions_white}, could be crucial in addressing complex classification challenges by leveraging discernible patterns within the feature space. This strategy could effectively overcome obstacles encountered in the delineation between classes of stars.

\subsection{Quantum Kernel Architecture and the Process Flow}

QKL leverages the principles of quantum computing to enhance classical machine learning frameworks. At the core of QKL is a VQC that acts as a quantum feature map, mapping classical data into an expanded quantum feature space. The parameters of the VQC are fine-tuned using classical optimization algorithms to improve the feature space representation.

\begin{figure}[ht]
\centering
\includegraphics[width=0.45\textwidth]{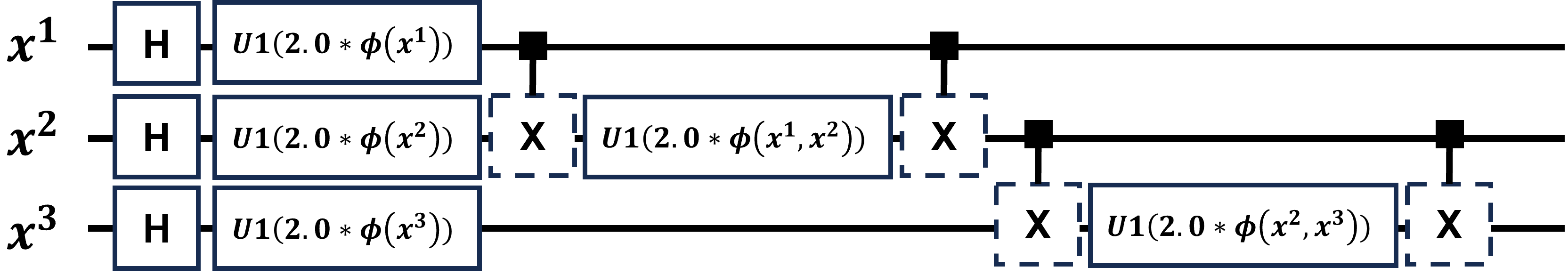}
\caption{Quantum Circuit of ZZFeatureMap, integral to the QKL algorithm for encoding data into quantum feature space.}
\label{fig:zz_feature_map}
\end{figure}

Within this framework, the ZZFeatureMap stands out as a specific implementation of a quantum feature map. It employs ZZ gates to intricately encode interactions between features onto quantum states. The quantum circuit is shown in Fig. \ref{fig:zz_feature_map}. This method is particularly adept at capturing complex correlations within the data, thereby improving the performance of machine learning tasks, such as classification, by utilizing the enriched feature space inherent to quantum systems.

QKL has the potential to surpass classical machine learning algorithms by leveraging these quantum-enhanced feature spaces. Moreover, when QKL is integrated with advanced computational resources, such as Quantum Processing Units (QPUs) or GPUs, it can significantly boost computational efficiency and learning efficacy.

\section{Results and Discussion}

\subsection{QKL for Large Stellar Dataset with Two Classes}

\begin{figure}[htbp]
  \centering
  \begin{tabular}{cc}
    \includegraphics[width=.45\linewidth]{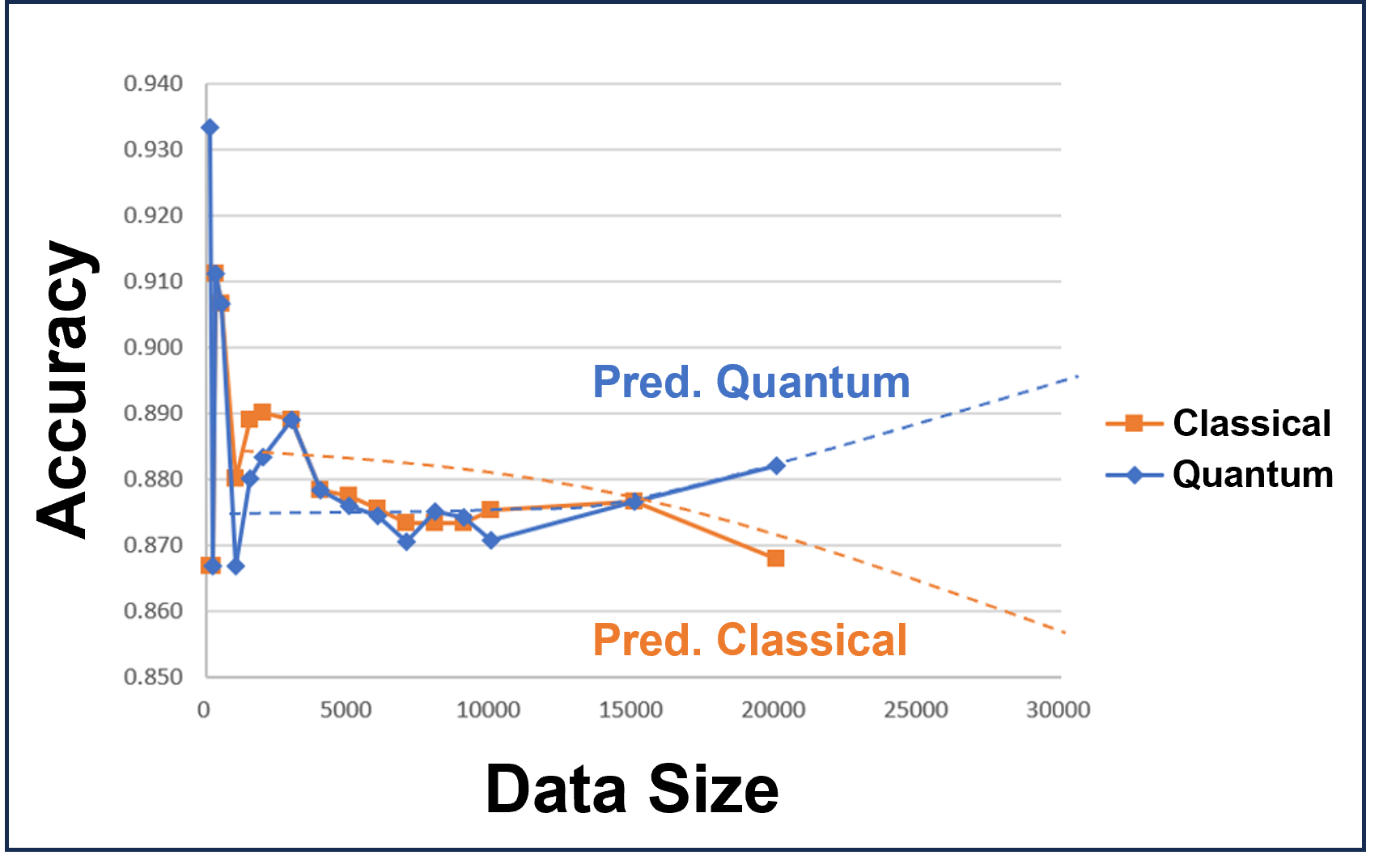} &
    \includegraphics[width=.45\linewidth]{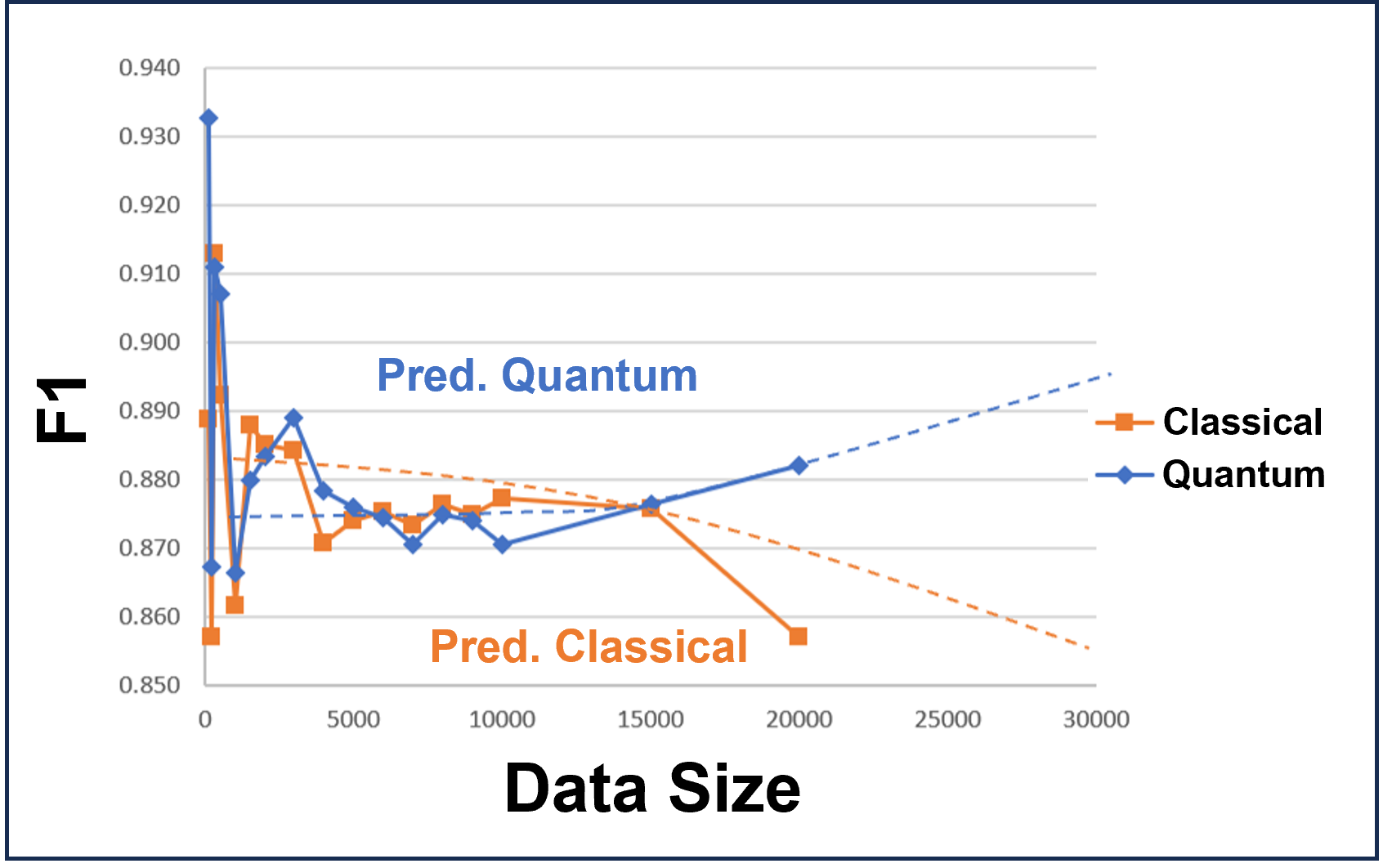} \\
    \includegraphics[width=.45\linewidth]{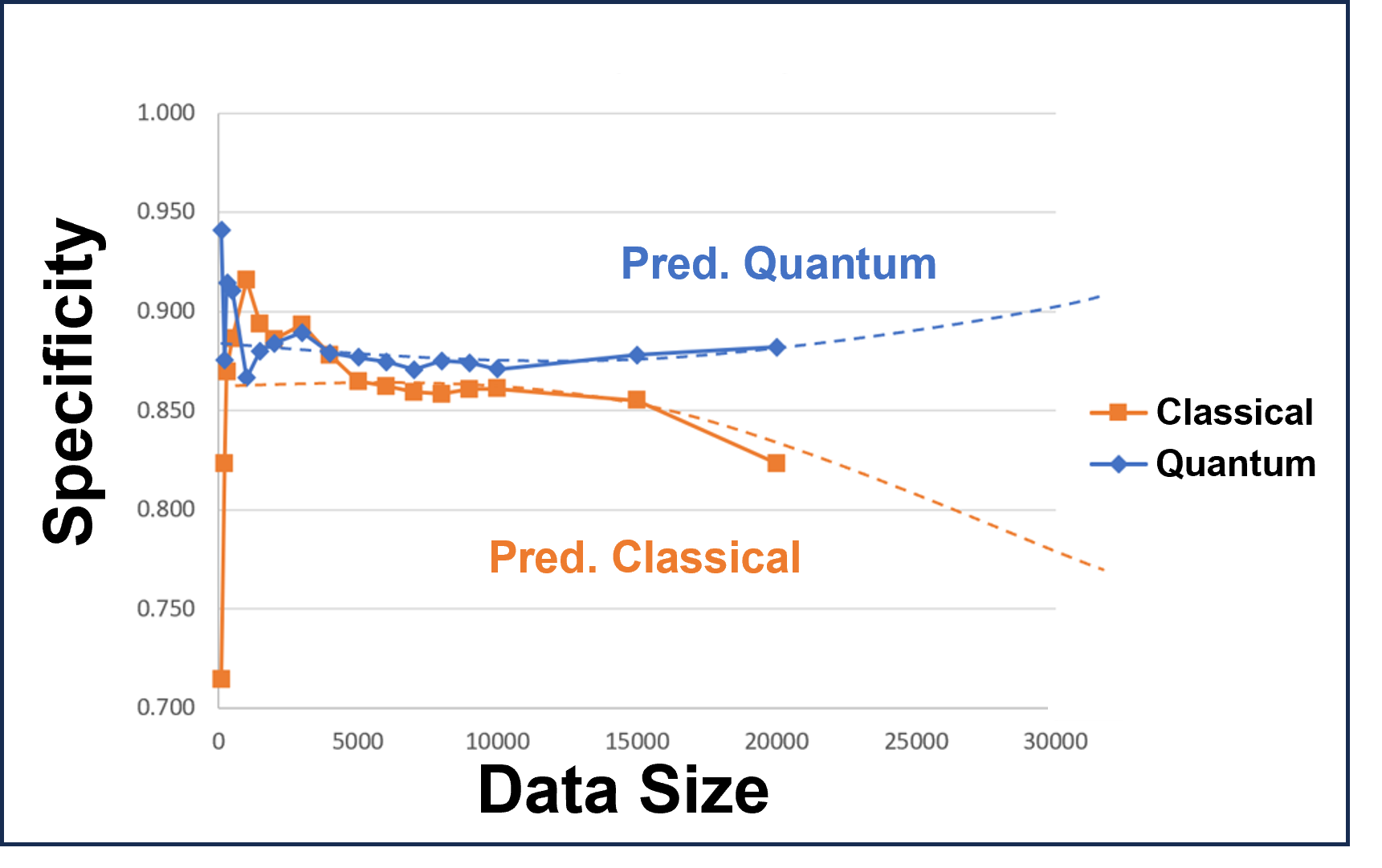} &
    \includegraphics[width=.45\linewidth]{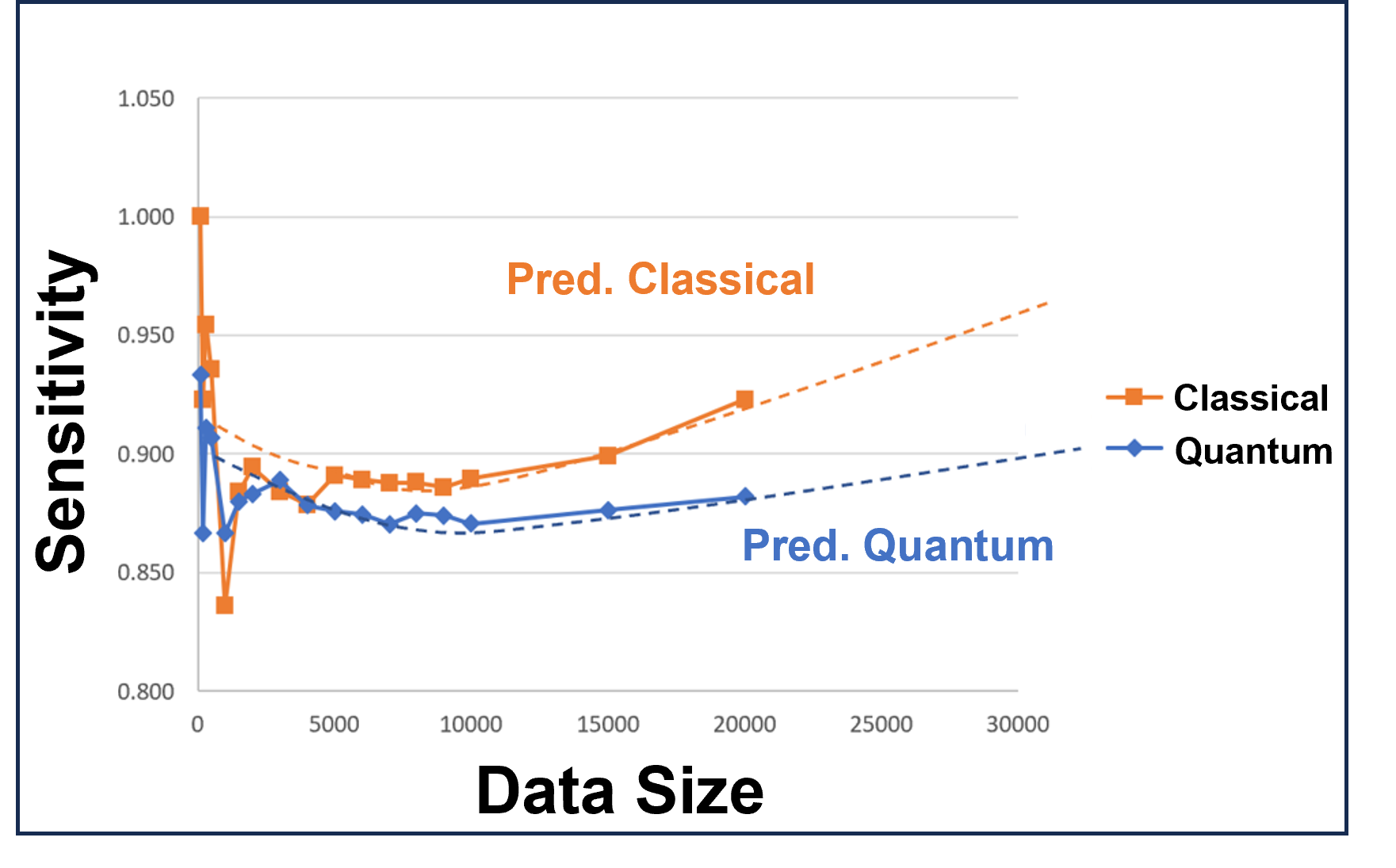} \\
  \end{tabular}
  \caption{Performance of QKL vs. Classical Algorithms. The graphs compare Accuracy, F1 Score, Specificity, and Sensitivity across varying data sizes for both QKL (Pred. Quantum) and classical predictions (Pred. Classical).}
  \label{fig:QKL_results}
\end{figure}

Our investigation into the application of QKL for star classification across large datasets is captured in Figure \ref{fig:QKL_results}. The performance metrics, including accuracy, F1 score, specificity, and sensitivity, indicate an enhancement in the QKL's predictive power with increasing training samples, especially beyond the 15,000-sample threshold.

As seen in the depicted results, QKL achieves superior accuracy, F1 score, and specificity when compared to its classical counterpart. Despite a marginal reduction in sensitivity, the quantum kernel upholds a consistent performance, maintaining a score near 0.9. This consistency underscores the robustness of QKL in classifying stellar objects. Moreover, the quantum kernel's performance demonstrates greater stability relative to the classical kernel across a variety of scenarios. This stability and improved performance are likely due to the quantum kernel's ability to process the complex features inherent in astronomical data more efficiently, resulting in enhanced prediction accuracy and stability.

\subsubsection{Classical Model for Benchmarking with Multi-class Classification}

\begin{figure}[ht]
  \centering
  \begin{minipage}[b]{0.45\linewidth}
    \includegraphics[width=\textwidth]{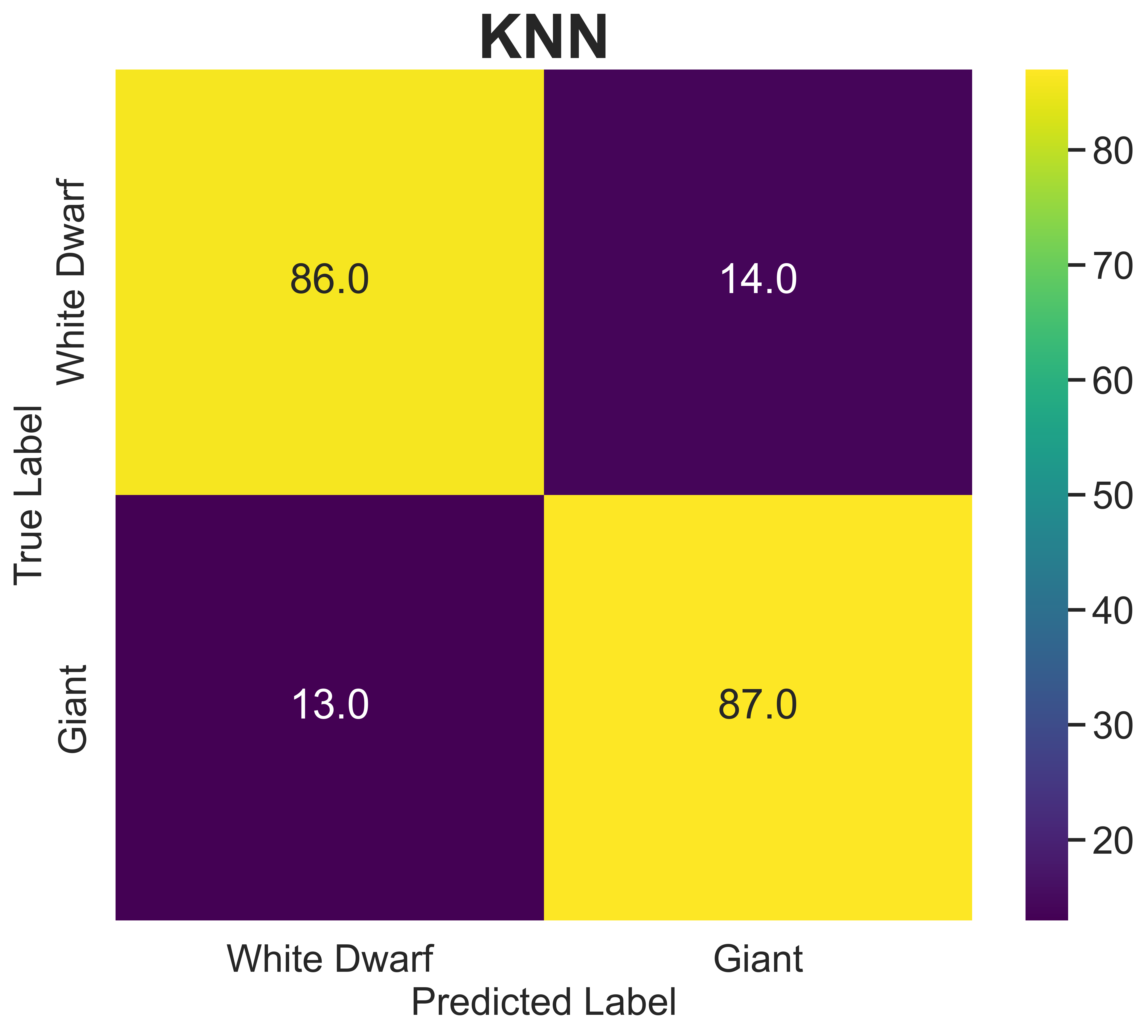}

  \end{minipage}
  \hspace{0.5cm} 
  \begin{minipage}[b]{0.45\linewidth}
    \includegraphics[width=\textwidth]{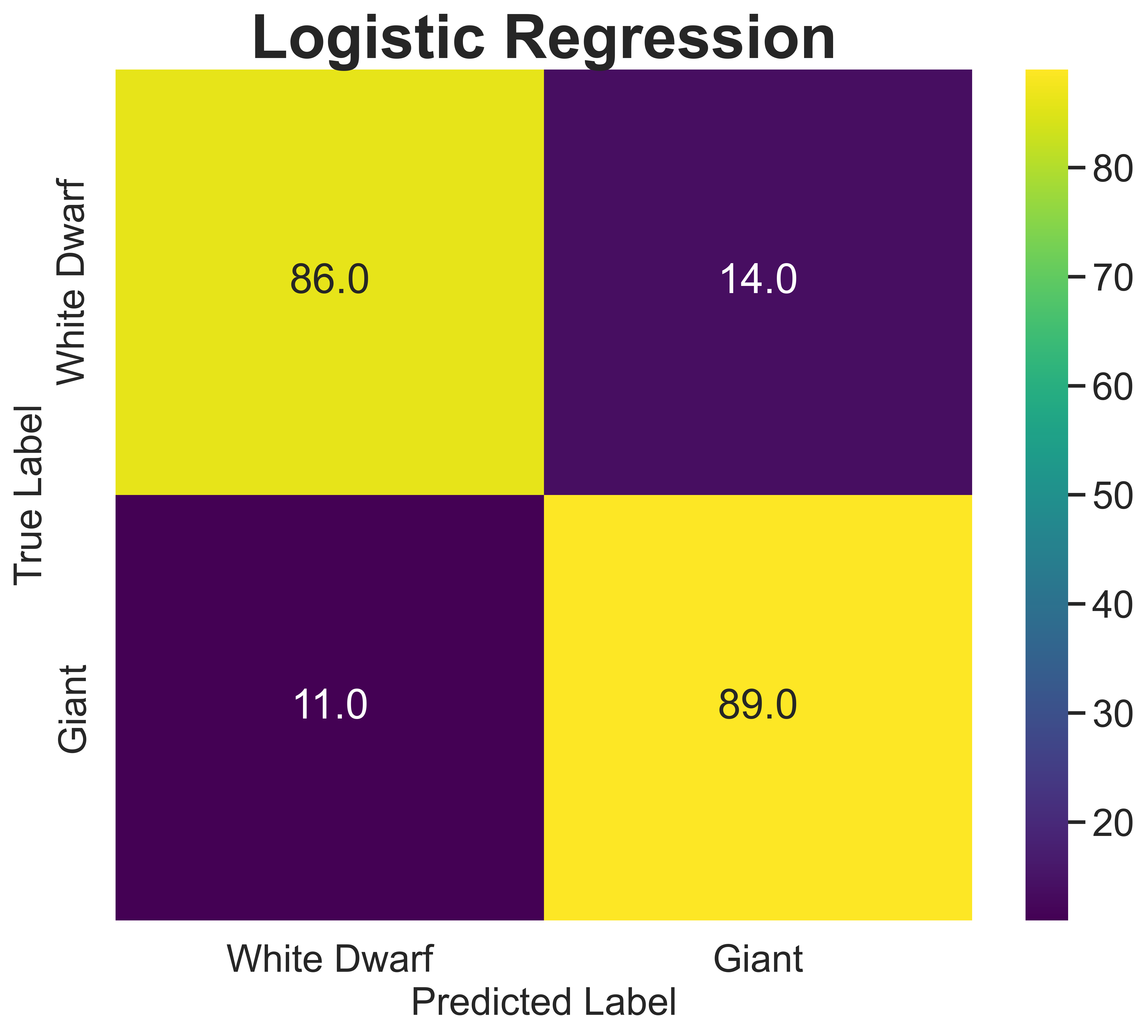}
  \end{minipage}
   \caption{Benchmarking the confusion matrix (expressed in percentages) for KNN and LR in binary classification.}
   \label{fig:confusion_matrix_qkl}
\end{figure}

In our benchmarking analysis, KNN and LR models exhibit commendable performance in binary classification, reaching an accuracy of approximately 0.86. Nevertheless, the complexity introduced by multi-class classification precipitates a decline in accuracy to 0.78 and 0.79 for KNN and LR, respectively, as shown in Fig. \ref{fig:confusion_matrix_qkl}.

\begin{figure}[ht]
  \centering
  \begin{minipage}[b]{0.45\linewidth}
    \includegraphics[width=\textwidth]{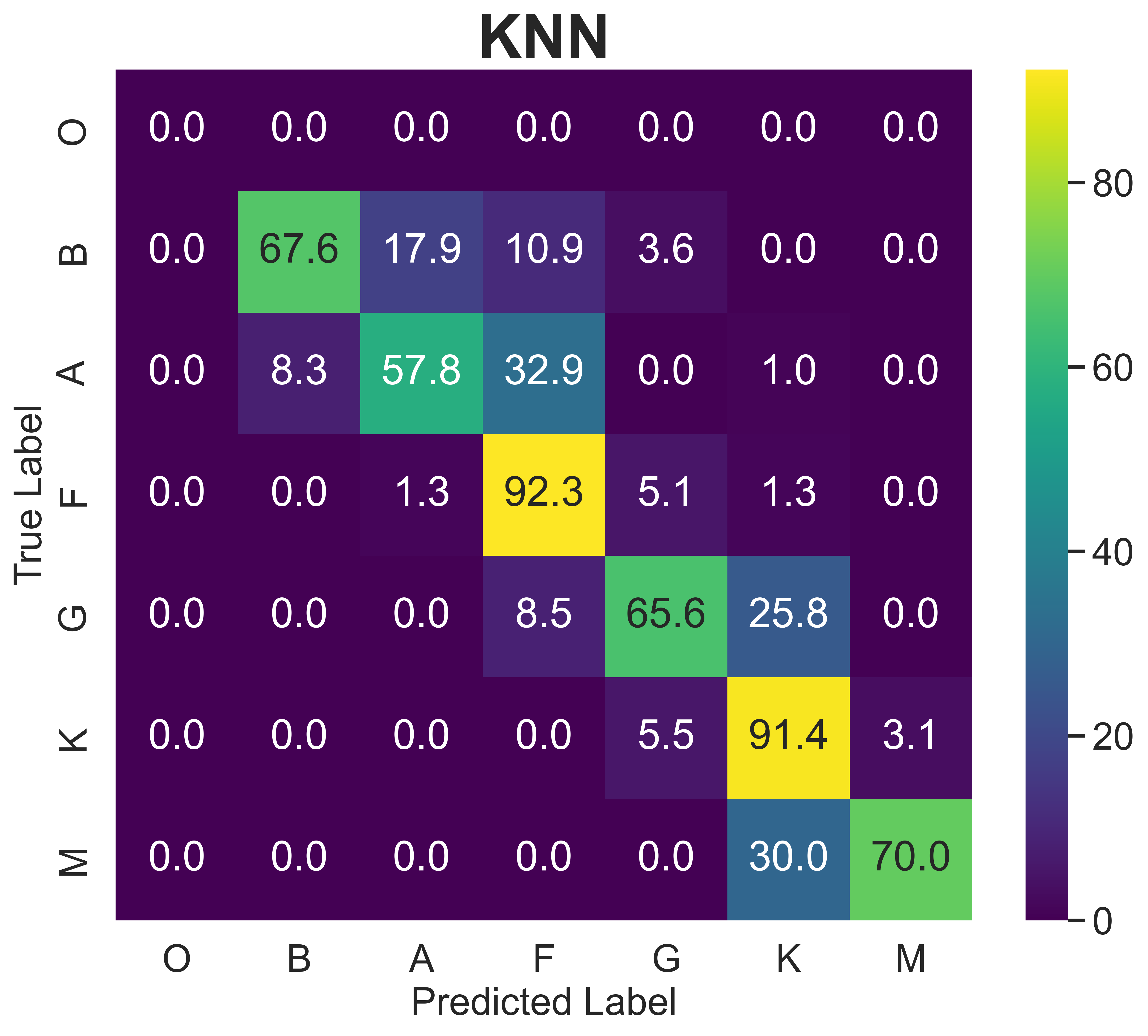}

  \end{minipage}
  \hspace{0.5cm} 
  \begin{minipage}[b]{0.45\linewidth}
    \includegraphics[width=\textwidth]{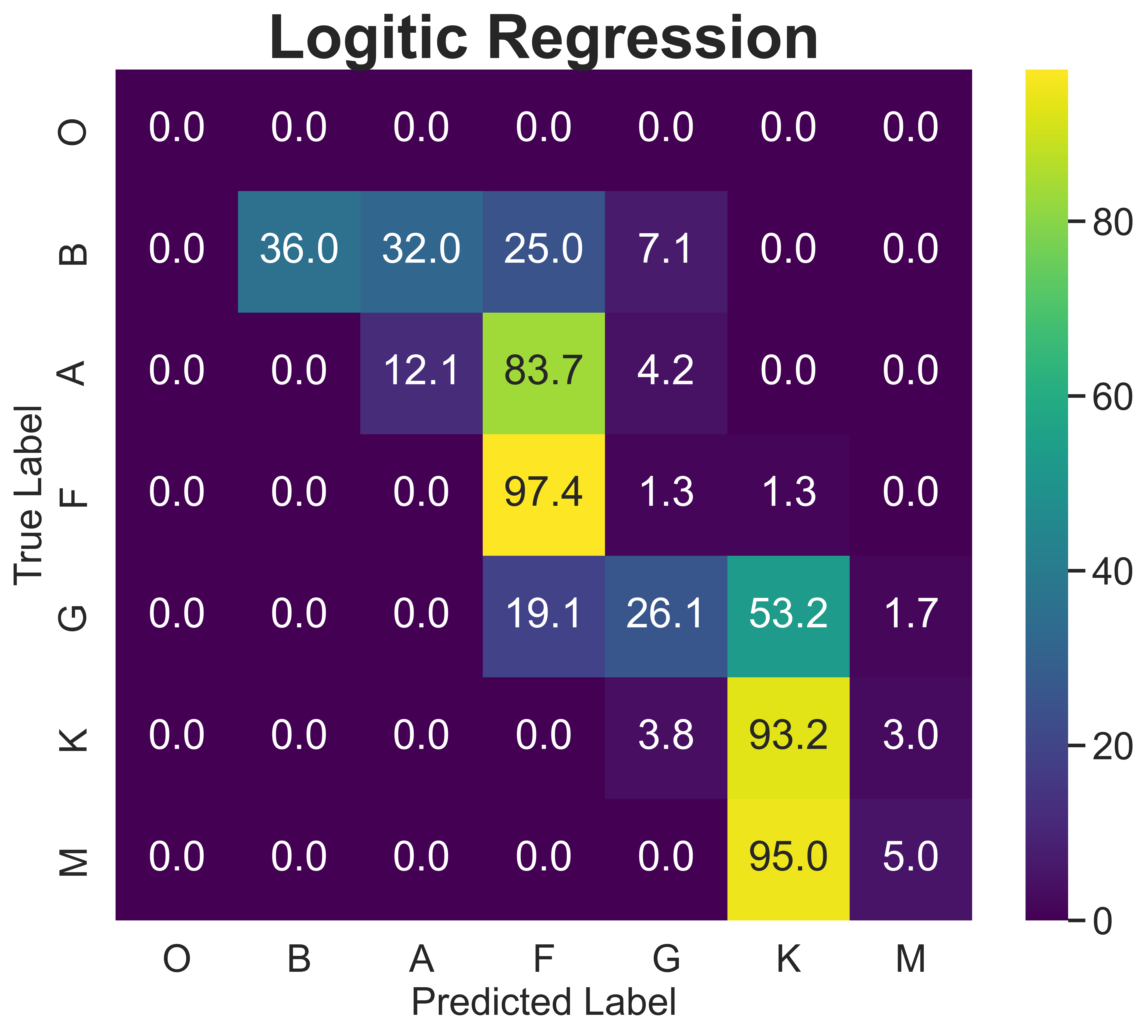}
  \end{minipage}
   \caption{Benchmarking the confusion matrix (as percentages) for KNN and LR in multi-class classification.}
   \label{fig:confusion_matrix_qkl_2}
\end{figure}

A noteworthy aspect of the findings is the difficulty in classifying stars of spectral types K and M as shown in Fig. \ref{fig:confusion_matrix_qkl_2}. These cooler, dimmer stars, coupled with their intricate spectra replete with absorption lines, pose a significant challenge to accurate spectral classification.

This complexity suggests that quantum machine learning approaches like QSVM could potentially enhance classification efficacy. The advanced computational capabilities of quantum methods may more adeptly navigate the intricate, high-dimensional data structures inherent in stellar classification, thereby refining the accuracy and efficiency of predictive models.

\subsubsection{Result of QKL for Multi-class Classification}
\begin{figure}[ht]
  \centering
  \begin{minipage}[b]{0.45\linewidth}
    \includegraphics[width=\textwidth]{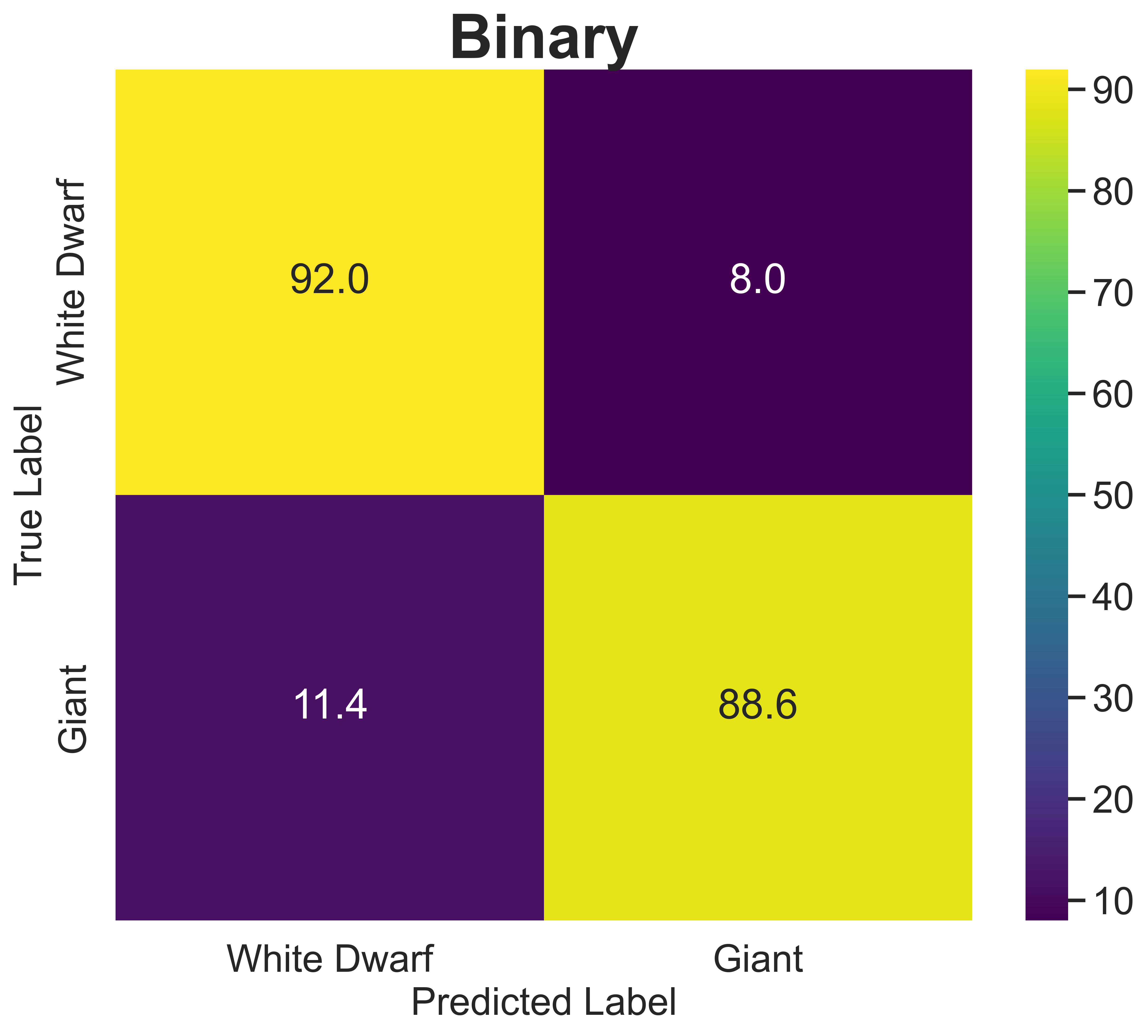}

  \end{minipage}
  \hspace{0.5cm} 
  \begin{minipage}[b]{0.45\linewidth}
    \includegraphics[width=\textwidth]{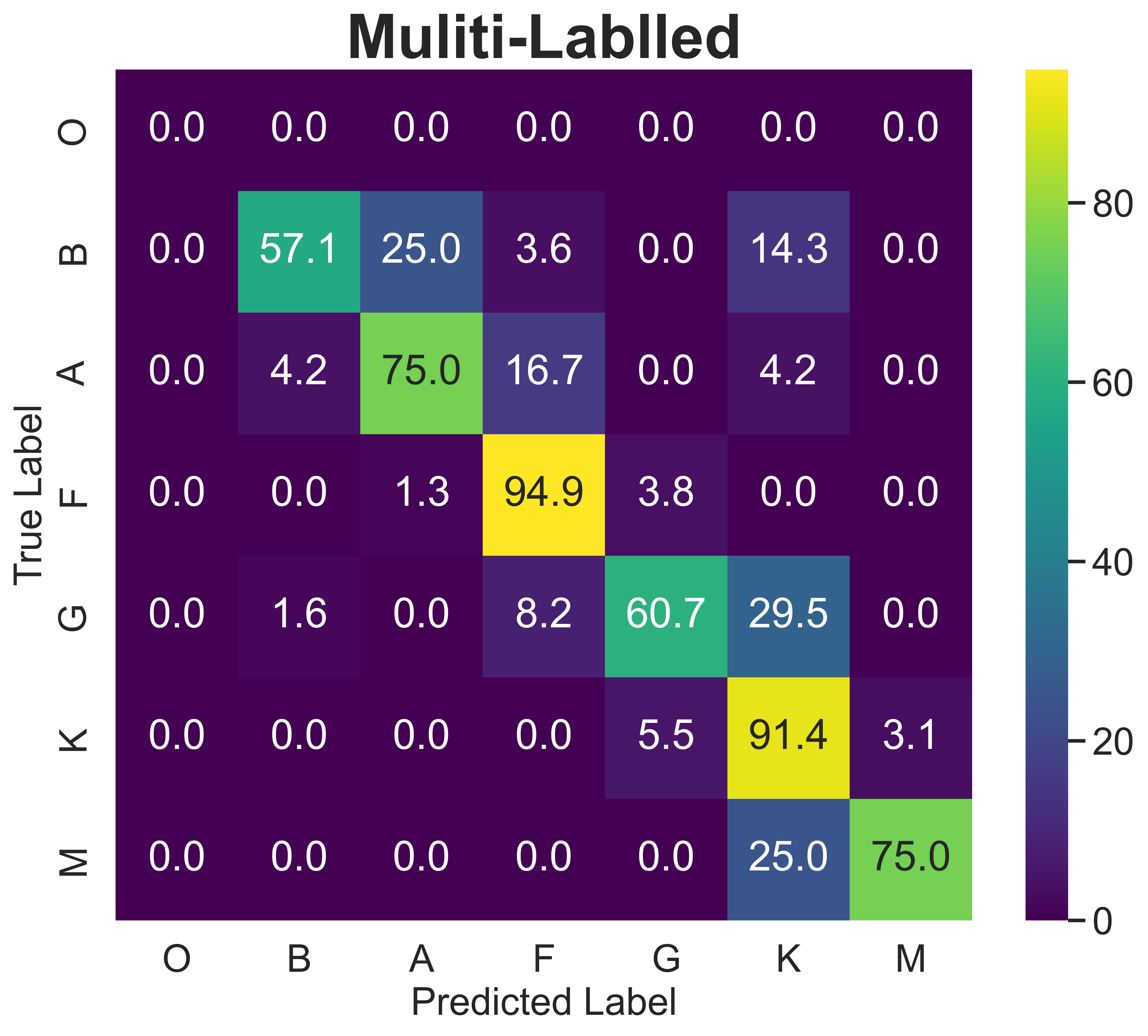}
  \end{minipage}
   \caption{Confusion matrix for the QKL algorithm with absolute values and percentages}
   \label{fig:confusion_matrix_qkl_result}
\end{figure}

Employing QKL in binary and multi-class classification yields an accuracy of about 89\% and 83\% accuracy respectively, surpassing classical benchmarks of around 5\%. Notably, in multi-class scenario, our quantum model demonstrates superior performance in classifying star types A \& F and K \& M, as visualized in Figure \ref{fig:confusion_matrix_qkl_result}, when compared to traditional methods such as KNN and LR. These findings underscore the potential of quantum machine learning to enhance the classification of stars, particularly for typically challenging classes.

\subsection{GPU Acceleration with cuQuantum for Quantum Kernel Encoding}

\begin{figure}[ht]
    \centering
    \includegraphics[width=0.45\textwidth]{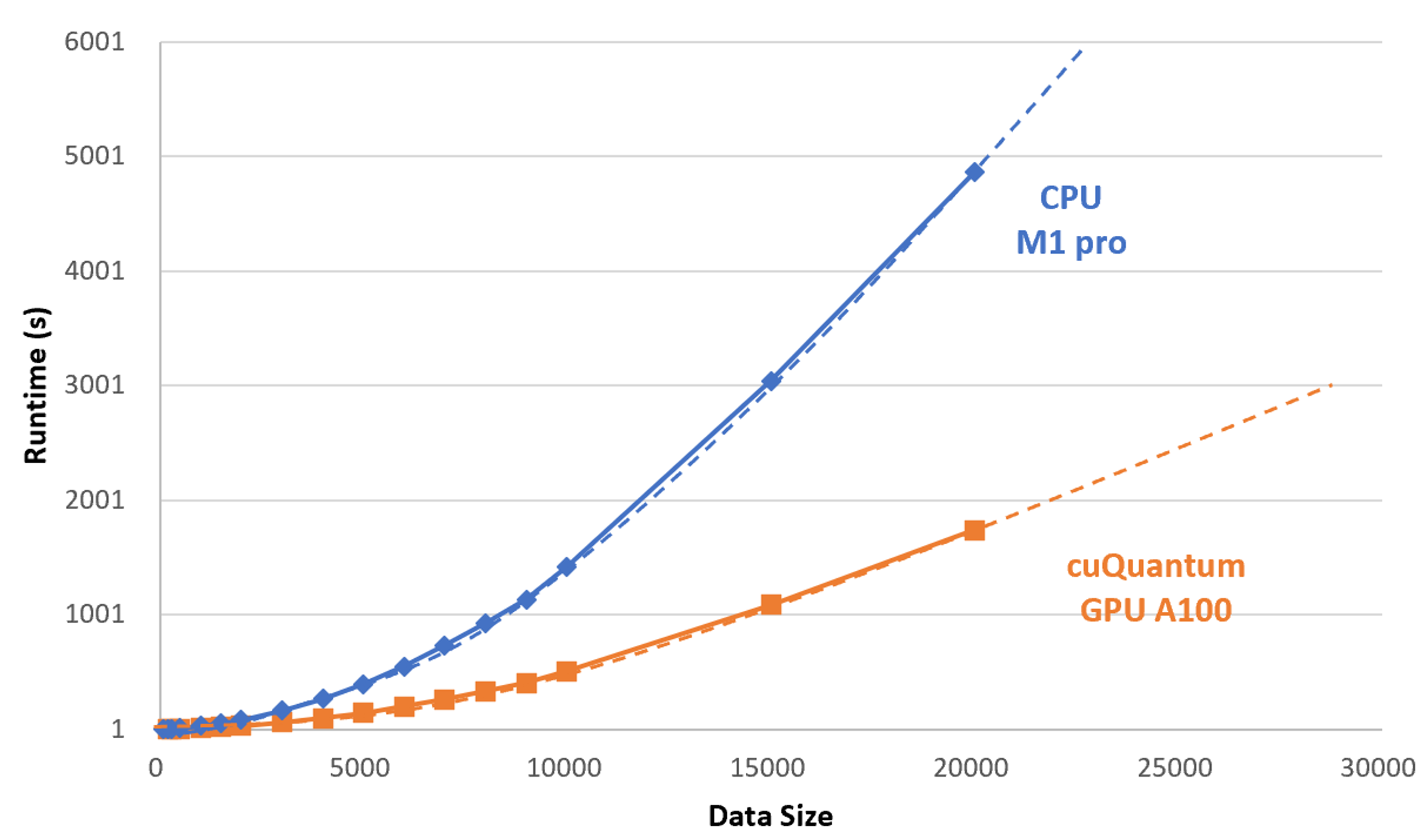}
    \caption{Benchmarking the QKL algorithm on an M1 Pro CPU and an A100 GPU with cuQuantum.}
    \label{fig:cuQuantum}
\end{figure}

Quantum computing holds the potential to solve complex problems in machine learning. However, training quantum models is computationally intensive and necessitates specialized hardware. QKL, a technique involving the pre-processing of data on classical hardware prior to training a quantum kernel, has demonstrated promising outcomes. CuQuantum, an NVIDIA software library, facilitates the training of quantum models on GPUs, thereby enhancing both the speed and accuracy of predictions. This acceleration stems from the GPUs' parallel processing capabilities. Moreover, the integration of CuQuantum with CUDA reduces computational costs, rendering QKL more accessible and scalable. Benchmarking results, as depicted in Fig. \ref{fig:cuQuantum}, compare the performance of the QKL algorithm on an M1 Pro CPU and an A100 GPU utilizing the CuQuantum library. These results showcase a significant speedup, with the GPU implementation outperforming the CPU by a factor of three, particularly when the QKL algorithm handles four features. This comparison underscores the efficiency and performance benefits of leveraging GPUs for quantum machine learning tasks.

\section{Conclusion}
Benchmarking studies have substantiated that QKL offers a substantial improvement over traditional machine learning techniques, achieving an accuracy of 83\% by leveraging effective quantum feature mapping. The integration of GPU acceleration, particularly through platforms like cuQuantum, has been shown to potentially triple the training speed of quantum kernels in comparison to conventional CPU-based methods. This enhancement not only mitigates computational costs but also expands the potential for broader application, as evidenced within the domain of stellar classification.

In the context of this paper we demonstrate stellar classification with QSVM whereas the coming application of quantum machine learning is not limited and open to broader studies on classifying other astronomical objects. The implications of QKL for the fields of AI and quantum computing are profound, suggesting a future where machine learning applications are revolutionized not only within the realm of stellar astrophysics but also in the broader spectrum of scientific inquiry. As these quantum computational approaches mature, they are poised to become pivotal tools in the advancement of AI, offering novel solutions to complex problems across various disciplines.

\section*{Data Availability}

The dataset analyzed during the current study is publicly available and was sourced from the Kaggle platform. The dataset, is titled ``Star Categorization - Giants and Dwarfs''. It can be accessed and downloaded directly from the following URL: \url{https://www.kaggle.com/datasets/vinesmsuic/star-categorization-giants-and-dwarfs}.

\section*{Acknowledgment}

The authors extend their gratitude to Tai-Yue Li for his invaluable discussions. K. C., X. X., and H. M. would like to express their gratitude to Xanadu for hosting QHack2023. This work is an extension of their QHack2023 winning project, ``Durchmusterung." The code implementation and HPC resources were supported by the NVidia cuQuantum team and the Run:ai platform. This research received support from the IBM Quantum Researchers Program. Special acknowledgment goes to QuantumPedia AI for their contribution. K.C. is also grateful for the financial support from the Turing Scheme through the Imperial Global Fellows Fund. H.-H. Chung is a member of the International Max-Planck Research School (IMPRS) for Astronomy and Astrophysics at the University of Bonn and Cologne.

\bibliography{mybib}

\end{document}